# Systolic Array Technique for Determining Common Approximate Substrings

Jacqueline E. Rice and Kenneth B. Kent

**Abstract**—A technique using a systolic array structure is proposed for solving the common approximate substring (CAS) problem. This approach extends the technique introduced in earlier work from the computation of the edit-distance between two strings to the more encompassing CAS problem. A comparison to existing work is given, and the technique presented is validated and analyzed based on simulations.

**Index Terms**—bioinformatics, FPGAs, reconfigurable hardware.

———————————— ◆ ————————————

## 1. INTRODUCTION

A common problem in bioinformatics is that of determining, within two or more strings of DNA, a common approximate substring (CAS) [1], [2]. A search for a CAS is considered to be successful if any common string of symbols is found within all of a given series of sequences, allowing for a certain amount of error. The goal is to investigate the use of field programmable gate array (FPGA) technology to combine the flexibility of software and the acceleration of hardware in finding a solution for the CAS problem.

This work presents a problem-specific systolic array implementation designed for FPGAs. FPGAs are the best choice for our design as the intent is that the initial DNA string will be preprocessed into a number of trees. Each node of these trees will then be implemented on a FPGA in a design specifically targetted for this DNA string. In other works this may be referred to as the database string. Many other strings (sometimes referred to as query strings) can then be compared in the search for one or more CASs. A preliminary version of this work was presented at ISCAS 2006 [3].

Although the primary contribution in this work is the proposed architecture, the work has been verified through simulation illustrating its feasibility. Results are reported in Section 4.1.

## 2. Background

The problem we wish to solve is that of determining which, if any, substrings within a given set of strings are common to all of the string sequences in a given set. The problem is made considerably more complex by allowing the incorporation of error factors in the matching process. These factors may include insertions, deletions or replacements of symbols within a substring. This is a technique used in DNA sequencing, where the discovery of a substring similar to that found in a known protein or family of proteins may provide information about the function of a newly sequenced gene [4]. The discovery of homologous sequences and families begins with the search for common substrings, also referred to as signals, or motifs [4], [5]. The goal, or challenge in this problem is to find similar sequences of symbols, where the length of the sequence or motif is a predefined value (m). We should note that the common string need not appear in the given set of strings, which is crucial in this

TGACTCGACC
TACTGCCTCG
CTGGCTAATA
ATTCCTGACT

Fig. 1. An example of common approximate substrings of length 5 with an error of 1. Solution motifs for this example are: TGACT, TGCCT, TGGCT, and TGACT.

area of work. The search space is a given set of n DNA strings, or sequences, also of a defined length (l). Finally, the allowable number of errors in the match between a given motif and a substring within a DNA string is set at d. In this work we limit our

———————————————

- *J. E. Rice is with the Department of Mathematics and Computer Science at the University of Lethbridge, Lethbridge, AB, Canada, T1K3M4.*
- *K. B. Kent is with the Faculty of Computer Science, University of New Brunswick, Fredericton, NB, Canada, E3B5A3.*

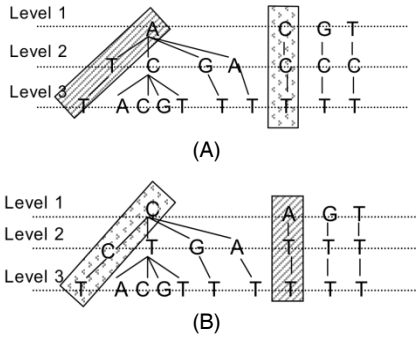

Fig. 2. An example illustrating how motifs ACT (A) and CTT (B) generate two distinct forests when d = 1. The shaded areas indicate where full trees, or complete branches of trees, can be shared.

(A)                    (B)

Fig. 3. (A) An example of how a level 2 node processes a character when the character does not match, and (B) an example of how the node processes a number.

allowable error factor to that of a simple replacement of one symbol with another; shifts and/or gaps in the sequence are not permitted, but will be considered in future work. Fig. 1 illustrates how a motif of length 5 can be found within 4 DNA strings, allowing an error of 1. We say that any common substring, allowing for the error factor, is a possible solution for the given sequence of strings. The reader is directed to [6] in particular Chapter 10 "Parallel Implementations of Local Sequence Alignment: Hardware and Software" and Chapter 28 "FPGA Computing in Modern Bioinformatics" for further details on DNA sequencing and FPGAs.

# 3   APPROACH

The first step in processing the strings from the database is to preprocess the first search string. It is necessary to partition the string into l−m+1 motifs, and then for each motif generate all possible motifs that are distance d errors from the generating motif. These are then stored as a forest of trees, where each tree represents all possible motifs for a given generator. Sharing of trees is possible in a restricted way. Entire paths from top node to leaf node can be shared, and identical upper portions of paths can be shared; however, unique paths must result in unique leaf nodes. An example is shown in Fig. 2. Note that in this example the trees are structurally equivalent, as they would be for any trees generated in this manner. Thus they can be used to identify any character string with an error of 1. This means that, provided the two parameters are maintained, we can avoid having to repeat the process of circuit re-creation, synthesis, place and route, and downloading when we wish to process new motifs and strings.

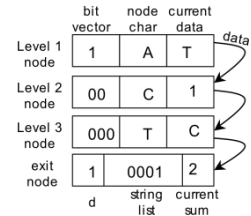

Fig. 4. An example of a path through a tree with its exit node recording the value of d, the current sum, and the 4 input strings for which this path has resulted in a potential CAS solution (currently just string 1).

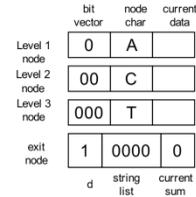

Fig. 5. The data initially stored in the tree nodes of a path representing motif ACT

## 3.1   Nodes

Each node in a tree has both processing and storage capabilities. The storage consists of the following:
• the character value of the node,
• a bit vector whose length is equal to the level on which the node resides, and
• a current piece of data. This may be a character or a numeric value as alternated in the input stream. The numeric value will never exceed 2 times the number of levels in the trees.

The character value of the node is required for performing comparisons with the characters that are streamed into the nodes. The bit vector is required to record the number of errors encountered at that node when performing comparisons; however, the memory of the bit vector (i.e. the number of errors to be remembered) is limited by the level in the tree at which the node resides. Root, or top nodes are at level 1 while the leaf or final nodes are at level m. The final piece of data is either a character for comparison to the node's own character or a numeric value that is used for summing the errors that have been encountered on the motifs' travel down through the levels of the tree.

Each node must be able to process either characters passed into it or numbers. If a character is passed to a node then the node

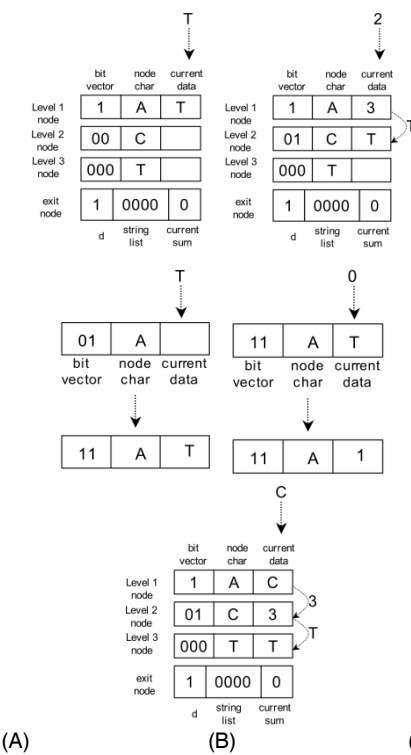

Fig. 6. (A)The first character to compare with, T, is passed into the top node. The characters do not match so a 1 is shifted into the bit vector.

(B) A 2 is next passed in; T moves to the next node and the value in the top bit vector is added to the 2 passed in. A 2 is passed in because we do not yet have a valid substring; until the length of the substring is >= m the sum values passed in begin at value d+1.

(C) The next character to compare with, C, is passed into the top node. The currently stored 3 is passed to level 2, and the T from level 2 goes to level 3; this is a match so a 0 goes into the level 3 bit vector.

must first perform a comparison of the given character to the node's own character. If their values match then a value of 0 is shifted into the right end of the node's bit vector. If their values do not match then a value of 1 is shifted into the bit vector. An example of this is shown in Fig. 3 (A). If a number is passed to a node then the node adds to it the value of the leftmost bit in the node's bit vector. An example of this is shown in Fig. 3 (B).

Most of the nodes in the system follow the requirements as given above. However below the leaf nodes, at level m + 1, there must also be a special type of node called an exit node. There is a one-to-one correspondence between each leaf node (at level m) and each exit node. Exit nodes collect and compare sums passed out of the above leaf nodes to d, and record which strings have found the path represented by that leaf node to be a potential CAS solution. If a sum value s is less than or equal to d then we record in the exit node the number of the string currently being processed in a bit vector. The reason for this is that any leaf nodes that result in any sums of d or less are satisfiable CAS solutions for the given string. Leaf nodes that result in potential CAS solutions for every string are verified CAS solutions for all strings. Once all strings have been processed through the systolic system we can determine which leaves are terminators for verified CAS solutions by checking which exit nodes have recorded a '1' bit for all input strings. Fig. 4 illustrates a path with its exit node and the data currently stored in it.

For example, if we begin with the length 3 motif ACT and allow 1 error then the clump of trees generated consists of 3 levels, with 10 leaf nodes as shown in Fig. 2 (A). Fig. 5 shows nodes on one path leading to a leaf node; this path represents the generating motif ACT. If we now begin processing an example string TCT then Figs 6 to 8 illustrate how the characters, alternating with digits, are propagated through the system. This example assumes there is only a total of 4 input strings.

The algorithm is as follows, assuming n strings of length l, and that we are searching for CAS motifs of length m with d permitted errors.

preprocessing step:
with first DNA string
for i=0 to l−m
      for motif consisting of characters i to i+m−1
            generate all possible motifs at distance d
            build tree consisting of those motifs
reduce nodes by merging trees with identical roots
processing step:

for  j = 2  to  n
for j=2 to n
  for k= 0 to l
(1)  input  character  k from  string  j  to  top  node(s)
    of tree array
  pass currently stored data to next node(s) down
  (2) if tick count ≤ m set x=d+1
  else set x=0
  input x to the top node(s) of the tree array
  pass  currently  stored  character  to  next
node(s)down
for k=l to l+m
 input – to top nodes in order to propagate final sums
to exit nodes
 determine which exit nodes are verified CAS solu-
tions

There are two types of node functionalities that are described below.

regular nodes:
if character data passed in
    if character matches node value
        shift 0 into right end of node's bit vector
    otherwise
        shift 1 into right end of node's bit vector
    if numeric data passed in
        take the number passed in and add to it the
leftmost  bit of node's bit vector

exit nodes:
if numeric data passed in
    if value is ≤ d
                set bit j representing string j in node's
bit vector to '1'
    if bit vector is all 1s then output '1'

## 4    Discussion

### 4.1 Analysis

Based on the size of data to be used we can draw the following conclusions.  There can be at most m levels in all trees, and at most

$$\sum_{i=0}^{\cdot} \cdot \binom{n}{,} \qquad (1)$$

leaf nodes in each forest.  This is the exact number of motifs generated for each group of m characters. For example, for m=10 and d=2 this results in 436 possible motifs. We are in this case limiting ourselves to 4 symbols in our alphabet (A, C, T, and G); further comments on this are given in Section 5 (Future Work).

Regardless of the number of leaves, the processing of each subsequent string is performed in constant time, requiring 2l + m steps for each string. This allows each character to move through the m levels of the forest as well as the intermediate numeric values for summing the errors encountered.

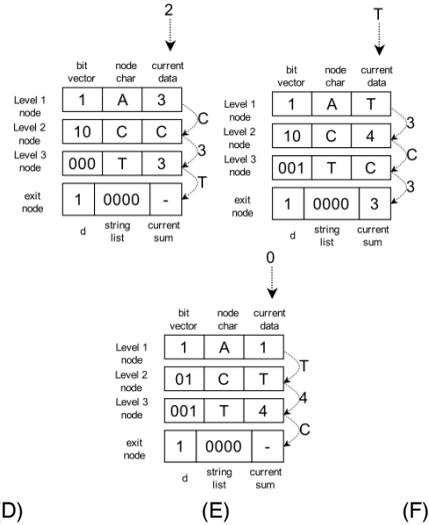

(D)                    (E)                    (F)

Fig. 7. (D) A 2 is next passed in, shuffling each of the pieces of data down to the next levels. Note that because the characters match a 0 is shifted into the rightmost bit of the level 2 node's bit vector.

(E) The final character T is passed in, causing the 3, C, and 3 values to move to the next levels down.  The middle 3 becomes a 4 as we add the leftmost bit of the bit vector at level 2.

(F) A 0 is ...

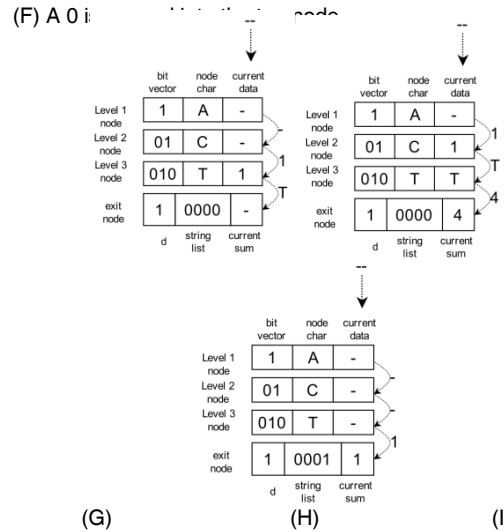

(G)                    (H)                    (I)

Fig. 8. In both (G) and (H) a - is passed in, in order to propagate the sum value for the entire motif down to the exit node.

(I) The final data is passed in, and the sum for the motif TCT reaches the exit node.  The sum is equal to d indicating that TCT is a potential CAS solution.

The big issue for any implementation of this type of technique is the memory limitations. Each regular node requires

• a maximum of m bits to store a bit vector,
• 2 bits for storing its character data, and
• memory for storing the numeric data (ef-
  fectively the count of non-matches previ-
  ously seen) to be passed into the node.

It should be noted that this memory does not need to be stored on chip; in fact it could be stored in external SDRAM.

If we allow 8 bits for the numeric data then the maximum sum value is 255, which should be far greater than will ever be required, as such a large value would mean that the motif lengths are 255, which would likely far exceed the memory capacity of any type of implementation. Exit nodes require

- a bit vector of length l to store the string list,
- memory for storing d, and
- memory to store the numeric data passed in as above.

It is likely that 4 bits would be sufficient to store d, as that would allow a maximum error factor of 15.

This is clearly a design intended for hardware implementation, as each node will in effect behave as a separate processor as the data is passed down through the structure on each clock pulse. The most important aspects to such a design are the node structure, as a great number of nodes are required, and the implementation of sharing of nodes. Optimization of the node structure has not yet been considered, as we have concentrated primarily on how sharing of the trees can be accomplished. Shared paths must be identified in the pre-processing step, which would take place in software. Finding sharing is accomplished by traversing a tree with the string to be added. When the tree "stops" in the middle, then one adds the remaining nodes. If the end is reached then the branch is complete. Once the hardware design is in place it is possible to compare to any number and any size of DNA strings desired. An area requiring some thought is that of careful selection of the first DNA string, the string used to generate the systolic array structure.

Initial implementation work targeting a Spartan 2E 200 FPGA found that a processor node implementation requires 8 CLBs and operates at 166.639 MHz. An exit node implementation requires 130 CLBs with a clock rate of 57.991 MHz. An implementation of the forest illustrated in Fig. 2 (A) (21 processing nodes and 10 exit nodes) resulted in 1452 CLBs operating at 57.991 MHz. However, since the exit nodes are only required to process every second input, the overall clock speed can be increased through simple clock management. A revised implementation of Fig. 2 (A) using a clock divider resulted in 1472 CLBs with a clock speed of 93.032 MHz.

## 4.2 Comparisons to Previous Work

Perhaps the earliest works related to this area would be [7], [8], [9] and [10]. The work in [9] introduces a hardware implementation of a systolic array technique published in [7]. It targets an application-specific reconfigurable chip referred to as SPLASH. Around the same time [8] introduced VLSI architectures for string and pattern matching; however this was before the advent of FPGAs and so the concept of a regular structure for this type of processing was, of itself, a novel concept.

Cheng and Fu [8] also introduced an algorithm for implementation on the new architecture. [10] presents the most recent of these implementations, each of which are focused primarily on computing the edit-distance between the reference string and one or more other strings. Our work removes a number of contraints that these earlier works are limited by, but loses the specificity of the results: that is, the location, within a particular error factor, of common motifs can be identified by our technique, but the edit distance costs for substitution, deletion and insertion are each fixed at 1.

For many years a basic local alignment search tool (BLAST) [4] has been commonly used; however to our knowledge the implementation of BLAST in [4] is strictly a software solution. Subsequent works such as [11] propose a number of algorithms for parallel string matching, but requires a specific type of reconfigurable mesh architecture for these algorithms. [12] introduces a bit-vector algorithm for approximate string-matching which uses dynamic programming, but again, assumes that a software implementation will be used. [13] discusses the FASTA3 program package, which offers a variety of algorithms implemented for string similarity searching, and later [14] and [15] use a hashing technique to index into portions of the DNA string and speed searches, but all implement only in software.

Since these earlier publications, other researchers have suggested the use of FPGAs in solving this problem. For instance, [16] suggests a dynamic programming technique implemented on a FPGA; however their approach requires run-time reconfiguration of the FPGA. In [17] another FPGA implementation of a systolic array technique is proposed, but this proposal also requires one or more reconfiguration phases before a solution can be found. [18] investigates a FPGA implemen-tation of the general case of the Smith-Waterman algorithm, which appears to be promising, but without any optimization of the programming elements or of the data storage, both of which are introduced in our work. As well, our work differs in that we design a solution beginning with a specific string to which we can then compare other strings. There is an upfront cost in our approach as opposed to a general approach such as [18], but it allows a more optimal design that we assume would allow the additional cost to be ammortized over repeated comparisons. Another FPGA implementation is suggested in [19]; they emulate portions of the well-known BLAST on a FPGA. Like our work, only a single pass is required for performing any comparison, and a tree structure allowing for parallel processing is utilized. Again, however, the intent of ours is to optimize for a specific problem, while this option is not factored into the work in [19].

Although it is very difficult to give a fair comparion amongst many of these works, a rough guide based on performance numbers reported in [19] is interesting. Table 1 gives the performance, in seconds, for two implementations reported on in that work, one hardware and one software. Like our solution their FPGA implementations also require some preprocessing time and this is not included in these numbers. Our numbers are generated by taking the observed achievable clock speed of our clock-divided implementation and using this to compute values based on the worst case number of steps that would be required for the given query length. We are not given a motif length to compare to, and so this value is left out of our equation; however as we can see in Table 1 a value of $m = 4$ or even $m = 100$ is not going to make a great deal of difference. As mentioned earlier, however, it is very difficult to perform a fair comparison; the reader is reminded that our technique does not, as yet, incorporate indels into the matching process, and is somewhat limited by the fact that we need a set motif length to search for. The system to which we are comparing is admittedly far more mature and complete.

One aspect that might be commented on is that of scoring. Many of the other described systems identify matching motifs with certain scores associated to the matches; these scores refer to the number of errors encountered in identifying the match. This is particularly important in many of the systems that work on heuristics and are not guaranteed to find the best matches. In our system we also have a scoring system inherent in the value of d that is both used to generate the forest of allowable motifs, as well as to keep track of whether a query string has a CAS solution. In our currently implementation we identify a CAS solution by identifying exit nodes that have sum values of d. A sum value of 0 might be achieved in some special case where the motif 'AAA' is matched with a query string of 'AAA', but we assume that this is unlikely to happen with any regularity. In most cases the sum value will be exactly d.

One might also assume that the size of the query might be an issue; however this is not the case for our technique. Instead, it is the size of the motif and the number of allowed errors that cause greater sizes (and numbers) of trees to be built. To the authors' knowledge, this is not necessarily information that is available when first searching for CASs amongst a database of strings. To overcome this problem, we might suggest computing an arbitrarily large motif size and error factor, such that

### TABLE 1
TIMES IN SEC. FOR PROCESSING VARYING LENGTH QUERIES USING METHODS ARE REPORTED IN [19].

| Platform | Query Length | | | |
|---|---|---|---|---|
| | 200 | 500 | 1000 | 2000 |
| Herbordt best FPGA | 2.5 | 50 | 10 | 20 |
| Herbordt BLAST software | 49±3 | 110±6 | 163±7 | 324±9 |
| our work (computed times) | .0000043 | .0000107 | .0000215 | .000043 |

all reasonable CAS solutions would be identified.

Unfortunately, one clear problem with this approach is the size requirements. Let us provide some numbers to demonstrate. One of the more recent FPGAs from Xilinx has approximately 50000 slices, which translates to about 13000 CLBs [20]. A motif of size m = 10 with allowable error factor d = 2 would result in at most 436 leaf nodes, as computed using Equation 1. Assuming no sharing of paths this would result in 4360 processor nodes and 436 exit nodes. At 8 CLBs per processor and 130 CLBs per exit node we end up requiring 34880 CLBs for processor nodes alone, and 56680 CLBs for the exit nodes. Clearly this has far outstripped the capability of the Xilinx Virtex 5 to which we previously referred! However, one must remember that we are using a form of decision diagram to share paths, and so there should be considerably fewer than 436 leaf nodes. The theory behind decision diagrams, and in particular ordered binary decision diagrams (OBDDs) gives us some ideas as to how we can reduce this number. In the usual type of ordered decision diagram, any two nodes with identical labeling and identical subtrees (i.e. trees below these nodes on the path from root to leaf) can be shared, thus removing one of the nodes and redirecting incoming paths from the removed node to the other. The reader is directed to [21] for further details on OBDDs. However in our application although we can share some nodes, we have more stringent sharing rules. For instance, in a "regular" decision diagram, there are only p leaf nodes where p is the size of the alphabet used (or in other words, the number of possible values the node may be assigned). In our application entire paths from top node to leaf node can be shared, and identical upper portions of paths can be shared; however, unique paths must result in unique leaf nodes.

A final comment on motifs may be relevant. We note that on the Web BLAST web page National Center for Biotechnology Information [22] information seems to indicate that common word sizes are 2 or 3, thus indicating the sufficiency of very small motifs.

## 5. Future Work

There are many aspects to DNA matching that we have not yet designed into this system. These include identifying reverses and repeats of motifs, possibly using algorithms such as those introduced in [23], [24] and [25].

We note also that on Web BLAST National Center for Biotechnology Information [22] the nucleic codes supported include A (adenosine), C (cytidine), G (guanine), and T (thymidine). However it goes on to allow N (any of A, G, C or T) and U (uridine), as well as a limited number of other degenerate nucleotide codes. It would not be a difficult matter to adjust our system to include these, although of course it would increase the size of the trees and forests to be implemented. For amino acid codes there is a considerably longer list of accepted codes; again, our system could be manipulated to manage this, but we have not yet attempted to do so and cannot comment on the impact it may have.

One possible technique to deal with indels might be to use a software wrapper to drive the matching process. We might record in a database, possibly using the onboard SRAM if available with the FPGA, where each motif is found. The preprocessing step must assign a unique id for each leaf node generated by a given motif, and then we can use a hash table to hash leaf ids to locations in the original string. Because our system performs so quickly we can then identify longer motifs with indels by using a software wrapper to do the following:

- search for motifs of size m, distance d;
- record all possible CAS solutions and their locations in the query string; and
- based on these CAS solutions match motifs that are a certain distance apart, resulting in motifs that may be found close to each other in the database string.

Another alternative might be to use a special character to encode insertions, such as - (as used on webBLAST). Deletions are more difficult, but a deletion in the query string can be considered to be an insertion in the database string, and so any system allowing for insertions in either the query or database string will cover this possibility as well.

## 6. CONCLUSION

This paper presents a design for a systolic type of structure intended for use in determining common approximate substrings amongst many DNA strings in a search set. This design is intended to use software for the preprocessing step, which will then generate a hardware description for implementation in a reconfigurable device. The implementation phase of this work is not yet completed, although preliminary results for each type of node are reported in Section 4.1. The authors have also been able to simulate an implementation for the forest of nodes shown in Fig. 2 (A), illustrating the feasibility of the design. Work in this area is continuing, and will ultimately result in a complete implementation which will be compared to previous work such as [26]. Results obtained thus far by this solution are positive.

There are a number of important differences between this and previous work. Many researchers suggest the use of a reconfigurable device in their solution but require multiple reconfiguration phases [16] or either processing in multi directions or a back-tracking phase [11], [26]. In our solution data progresses in one direction through the structure, thus giving us a fixed and constant time for determining CAS solutions. This work is also intended for use on an off-the-shelf FPGA, rather than a hybrid grid structure or any other specialized type of hardware.

A very important result of this work is that, in comparison to other techniques, this is the first and only technique to suggest the building of a forest in which portions of the component trees can be shared. As introduced by Bryant [21] structures referred to as Reduced Ordered Binary Decision Diagrams have come to be widely used in many areas due to the efficiency of applying operations on them and to the fact that they provide a unique representation for any given binary function. It is upon this concept that the notion of our forest of trees is built, and we anticipate that our work may be able to leverage many of the advances from recent studies into decision diagrams. We should note that [24] discusses a weighted suffix tree in this context, but to our knowledge this has not been used in a hardware implementation.

In addition to the above contributions, we also note that we did not find any other work in the literature that attempted to design a solution optimal to a particular problem instance, with a view to amortizing the extra cost over multiple uses. Given the intended use of our design; that is, to be used in finding common approximate substrings in multiple DNA sequences, we feel that this approach is likely to have a lower overall cost than many general approaches.

While our initial proof-of-concept testing has shown very favourable results, future work must involve further implementation and thorough testing. We hope to incorporate other algorithms that include considering repeats of motifs and reverses of the motif(s) in question. We also plan to incorporate other error factors in determining common motifs, such as gaps and deletions, into this work.

## 7. ACKNOWLEDGEMENT

This work was supported in part by a grant from the Natural Sciences and Engineering Research Council of Canada. Software and hardware used in this work was provided by the Canadian Microelectronics Corporation (CMC) .

**J. E. Rice** is currently an Associate Professor with the Department of Mathematics and Computer Science at the University of Lethbridge.  She is a member of the IEEE and the IEEE Computer Society.  Her research interests include reversible logic, representations of Boolean functions, and multiple-valued logic, as well as FPGAs and reconfigurable hardware.

**K. B. Kent** is currently an Associate Professor with the Faculty of Computer Science at the University of New Brunswick.  He is a member of the IEEE and the IEEE Computer Society.  His research interests include Hardware/Software Co-Design, Reconfigurable Computing, Software Engineering, and Embedded Systems